\documentclass[
    ,final            
  ]
  {aipproc}

\layoutstyle{6x9}

\begin{document}

\title{CAIXA: a Catalogue of AGN In the XMM-\textit{Newton} Archive - Correlations}

\classification{98.54.Cm}
\keywords      {Galaxies: active - Galaxies: Seyfert - quasars: general - X-rays: general}

\author{Stefano Bianchi}{
  address={Dipartimento di Fisica, Universit\`a degli Studi Roma Tre, via della Vasca Navale 84, 00146 Roma, Italy}
}

\author{Nuria Fonseca Bonilla}{
  address={XMM-Newton Science Operations Center, European Space Astronomy Center, ESA, Apartado 50727, E-28080 Madrid, Spain}
}

\author{Matteo Guainazzi}{
  address={XMM-Newton Science Operations Center, European Space Astronomy Center, ESA, Apartado 50727, E-28080 Madrid, Spain}
}

\author{Giorgio Matt}{
  address={Dipartimento di Fisica, Universit\`a degli Studi Roma Tre, via della Vasca Navale 84, 00146 Roma, Italy}
}

\author{Gabriele Ponti}{
  address={Laboratoire APC, UMR 7164, 10 rue A. Domon et L. Duquet, 75205 Paris, France}
}

\begin{abstract}
We presented CAIXA, a Catalogue of AGN in the XMM-Newton Archive, in \citet{bianchi09}. It consists of all the radio-quiet X-ray unobscured (N$_\mathrm{H}<2\times10^{22}$ cm$^{−2}$) active galactic nuclei
(AGN) observed by XMM-\textit{Newton} in targeted observations, whose data are public as of March 2007. With its
156 sources, this is the largest catalogue of high signal-to-noise X-ray spectra of AGN. All the EPIC pn
spectra of the sources in CAIXA were extracted homogeneously and a baseline model was applied in order to
derive their basic X-ray properties. These data are complemented by multiwavelength data found in the
literature: Black Hole masses, Full Width Half Maximum (FWHM) of H$\beta$, radio and optical fluxes. A systematic
search for correlations between the X-ray spectral properties and the multiwavelength data was performed for
the sources in CAIXA. We discuss here some of the significant ($>99.9\%$ confidence level) correlations.
\end{abstract}

\maketitle


\section{Correlations with radio emission and its origin in radio-quiet AGN}

\begin{figure}
  \includegraphics[height=.32\textheight]{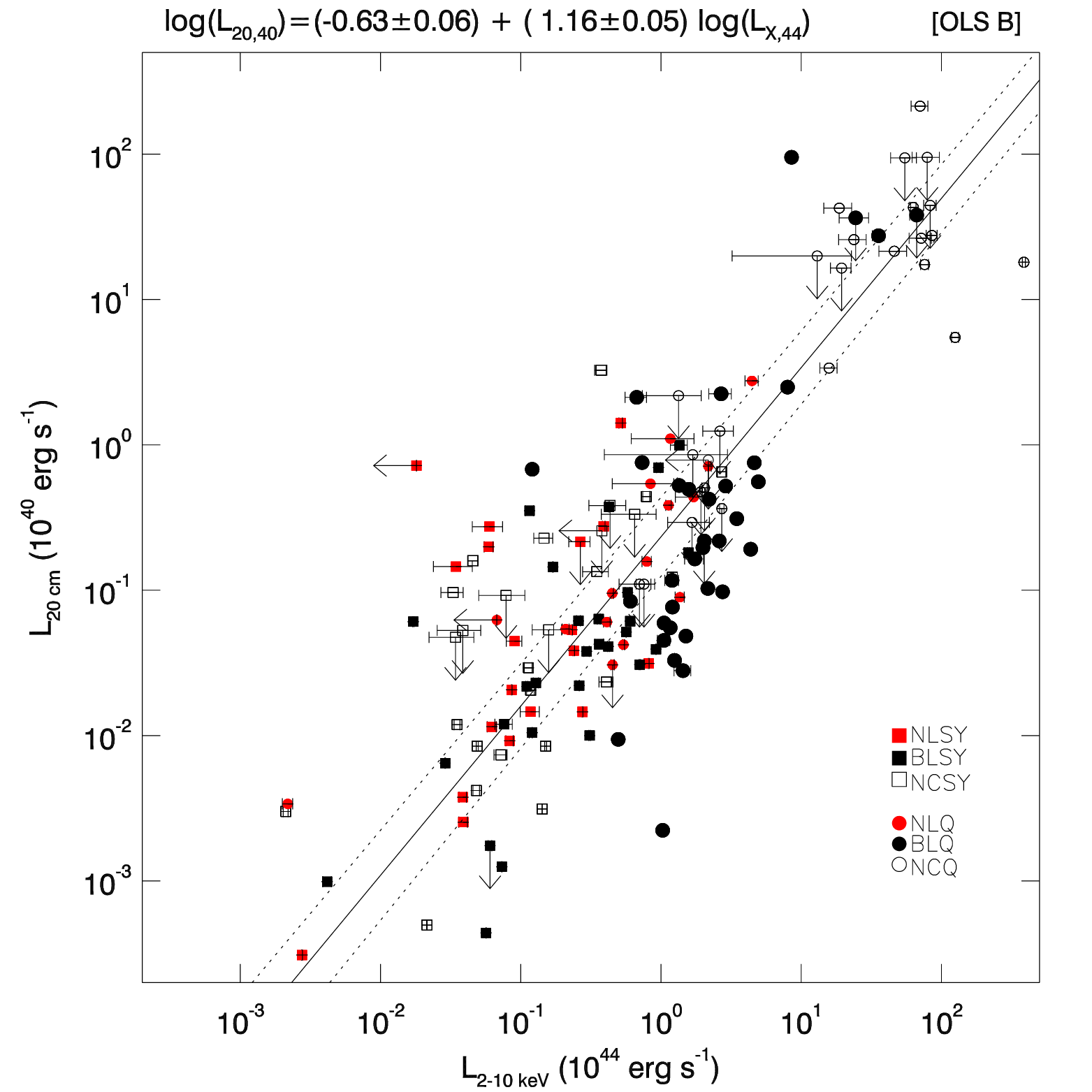}
  \includegraphics[height=.32\textheight]{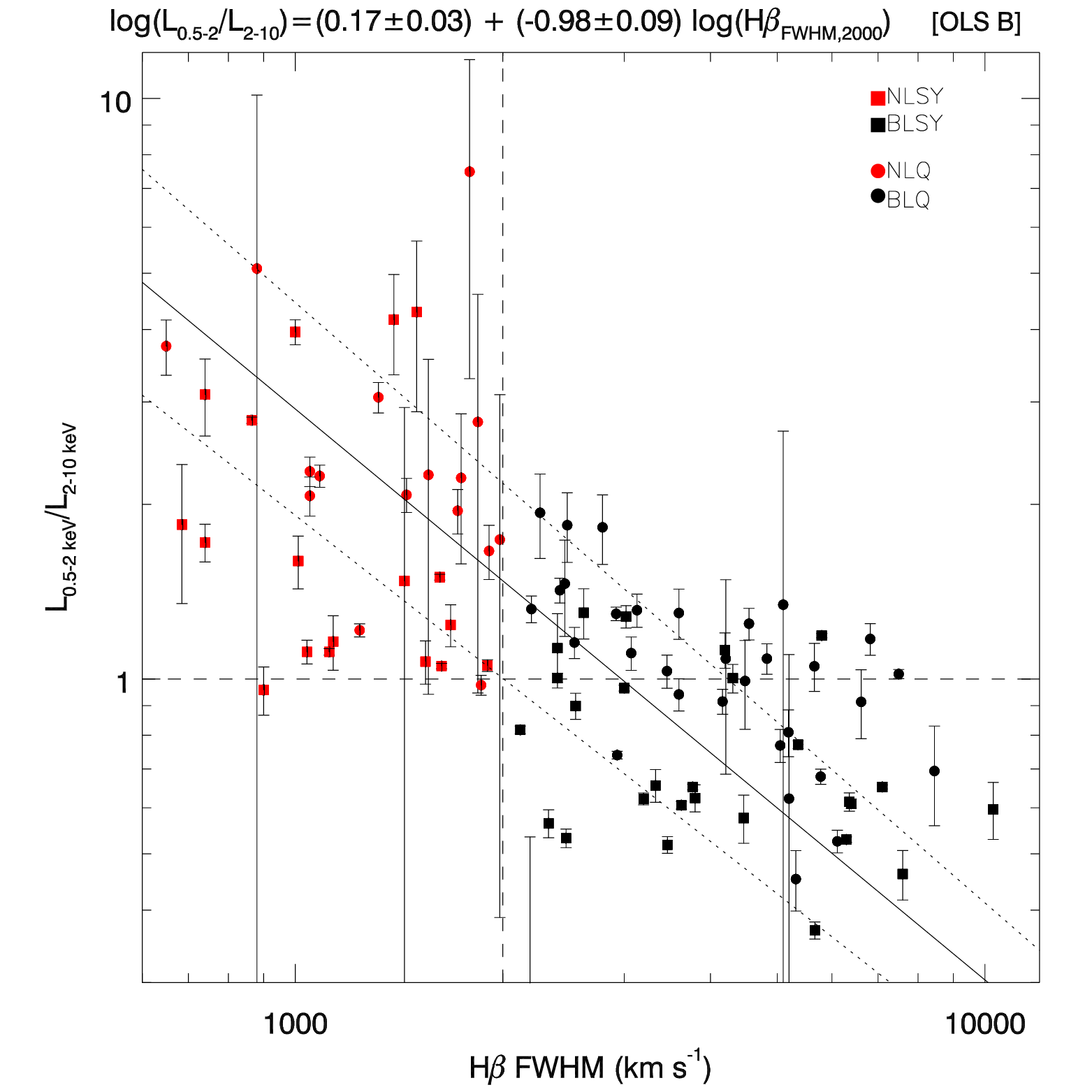}
  \caption{\label{lx_lradio}\textit{Left}: 2-10 keV vs 20 cm luminosity. The analytical expression for the best fit is on the top. \textit{Right}: 0.5-2/2-10 keV luminosity ratio vs. H$\beta$ FWHM. The anti-correlation between the two parameters is very significant. The different symbols refers to the classification of the objects, on the basis of their absolute magnitude and H$\beta$ FWHM: \textit{NLSY}, narrow-line Seyfert 1; \textit{BLSY}, broad-line Seyfert 1; \textit{NCSY}, not-classified Seyfert 1 (no H$\beta$ FWHM measure available); \textit{NLQ}, narrow-line quasar; \textit{BLQ}, broad-line quasar; \textit{NCQ}, not-classified quasar (no H$\beta$ FWHM measure available).}
\end{figure}

Very strong correlations are found between the hard (or soft) X-ray luminosity and the radio luminosity (at 6 or 20 cm): see Fig. \ref{lx_lradio}. Although the correlations are still very strong even when the effect of distance is taken into account via a partial Kendall $\tau$ test, a ``scrambling test'' based on $100\,000$ simulated datasets suggests that the role of distance may be dominant. Therefore, we decided conservatively not to derive any physical interpretations from these correlations. Although this conclusion applies only to CAIXA, we point out once more that care must be taken in assessing the real significance of any luminosity-luminosity correlation and in extracting physical information from its functional relation. We do not find any significant anti-correlation between the radio-loudness and the Eddington ratio, but this was expected since CAIXA does not include radio-loud objects.


\section{Correlations with H$\beta$ FWHM and the nature of the soft excess}

\begin{figure}
  \includegraphics[height=.22\textheight]{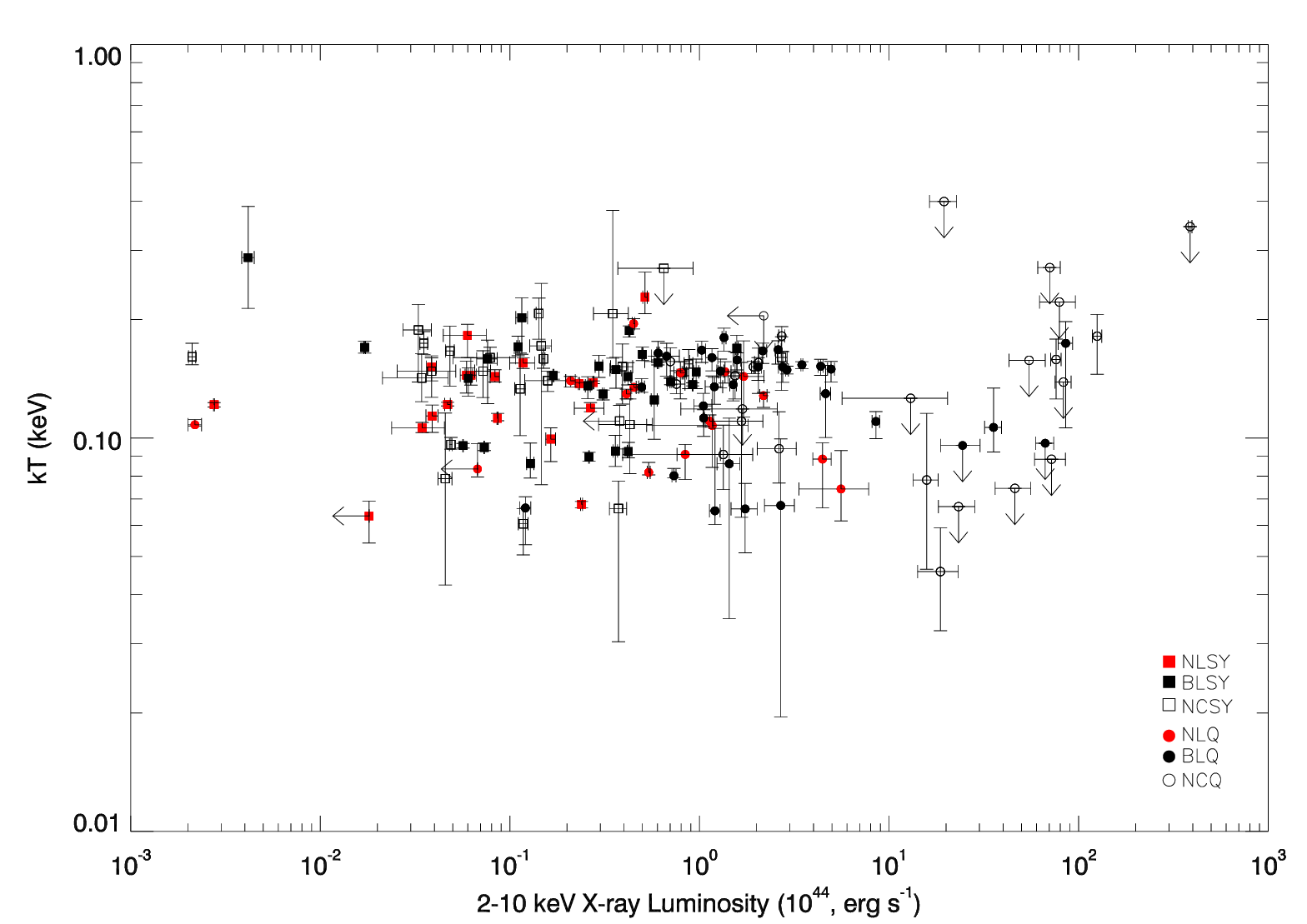}
  \includegraphics[height=.22\textheight]{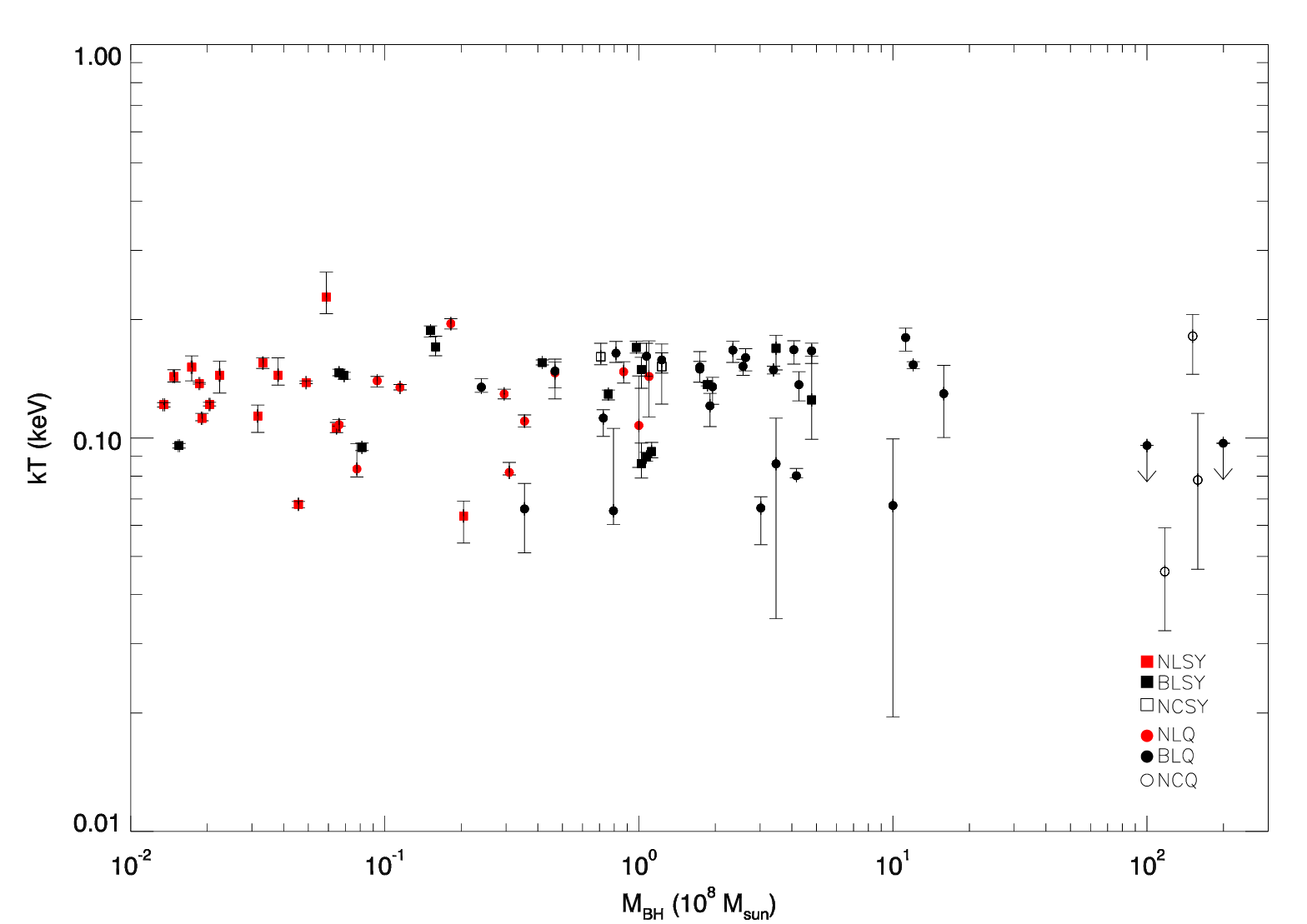}
  \caption{\label{bbkt}Temperature of the black body model, when used to model the soft excess in the objects of CAIXA. It appears completely unrelated to the X-ray luminosity (\textit{left}) and the BH mass (\textit{right}).}
\end{figure}

A very strong anti-correlation between the FWHM of H$\beta$ and the ratio between the soft and the hard X-ray luminosity is present (see Fig. \ref{lx_lradio}). In particular, our catalogue does not contain narrow-line objects with a $L_{0.5-2}/L_{2-10}$ ratio lower than 1. With respect to the correlation between the H$\beta$ FWHM and the soft X-ray slope (which, incidentally, is not significant in CAIXA) found in other works, we believe that our anti-correlation between the X-ray luminosity ratio and H$\beta$ is a more significant physical correlation between two \textit{a priori} independent properties of radio-quiet AGN, being model-independent. The interpretation of the soft excess as direct emission from the accretion disc was put in serious difficulty by several studies, which showed a `universal' temperature, independent on luminosity or BH mass \citep[e.g.][]{gd04,crummy06}. We confirm this result in CAIXA: see Fig. \ref{bbkt}.

\section{Correlations with BH mass and the X-ray bolometric correction}

\begin{figure}
  \includegraphics[height=.32\textheight]{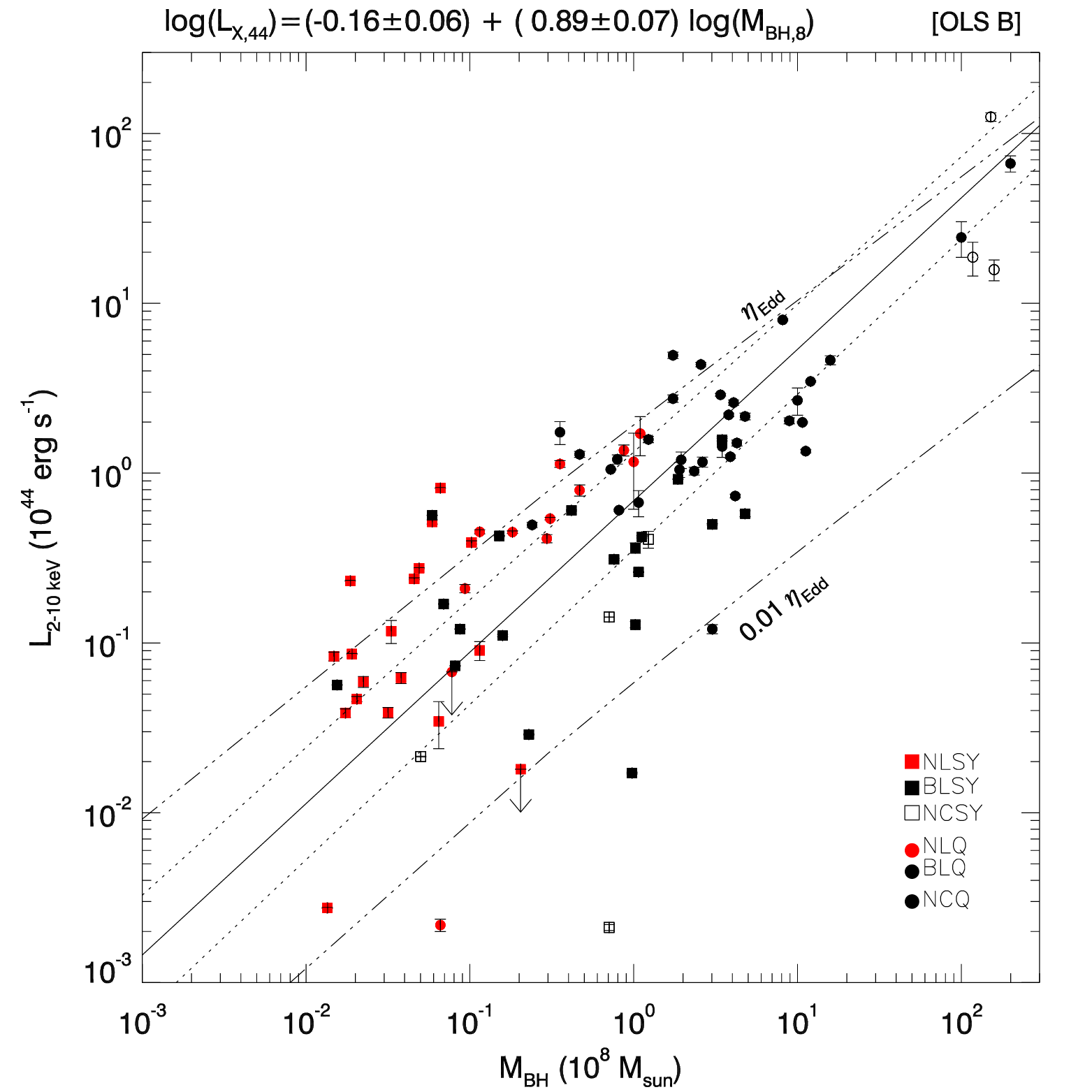}
  \includegraphics[height=.32\textheight]{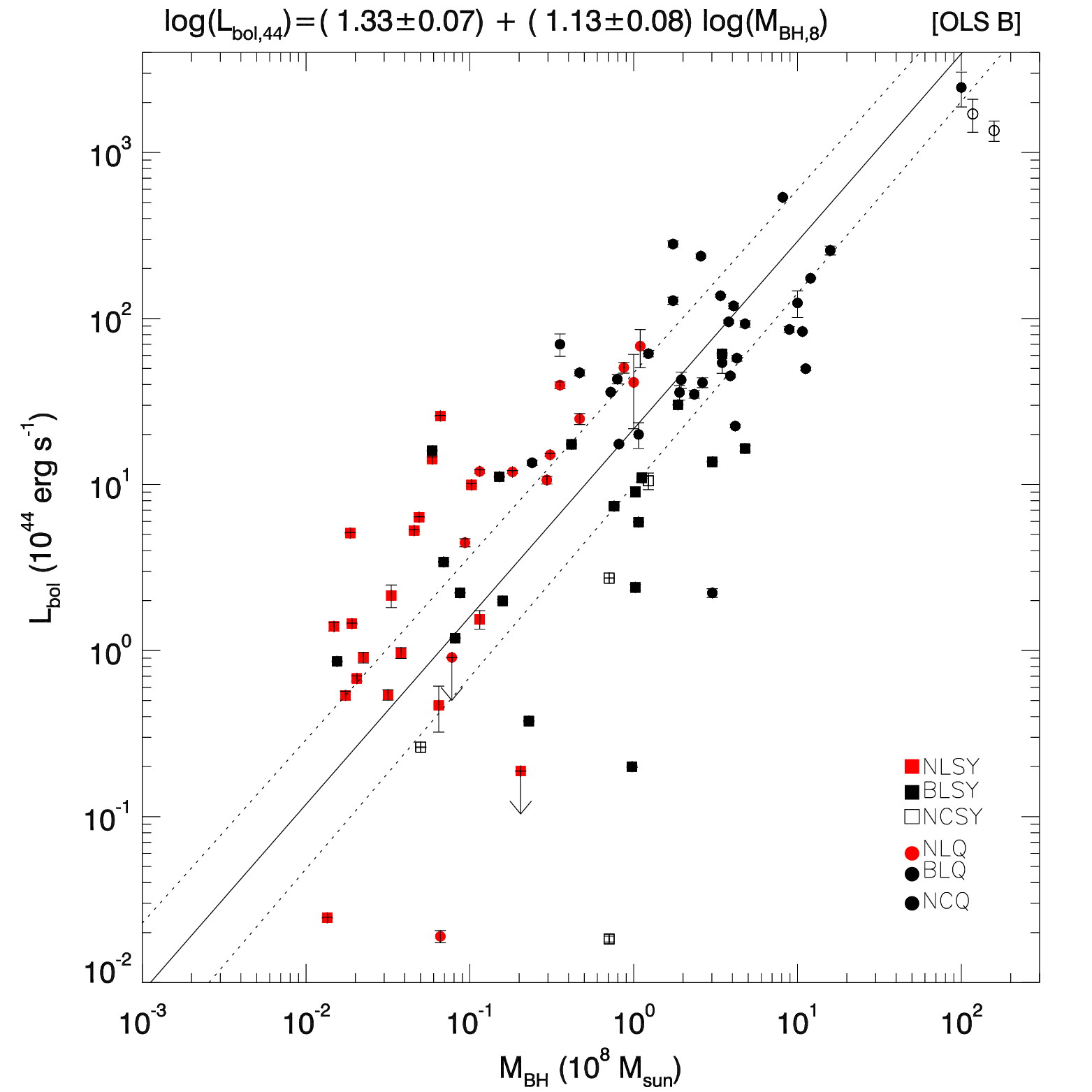}
  \caption{\label{lxmbh}\textit{Left} 2-10 keV X-ray vs. BH mass for all the sources in the catalogue with an estimate of the latter. The curves referring to Eddington rates of 1 and 0.01 are also shown. \textit{Right} Bolometric luminosity (luminosity-dependent correction) vs. BH mass for all the sources in the catalogue with an estimate of the latter. A linear regression fit is superimposed on the plots.}
\end{figure}

The strong correlation between the X-ray luminosity and the BH mass has a slope flatter than 1, suggesting that high-luminosity objects may be X-ray weaker (see Fig. \ref{lxmbh}). A luminosity dependent bolometric factor could explain this result, but the one proposed by \citet{mar04} seems too strong at high luminosities. In any case, a linear relation is recovered when using radio luminosity, suggesting that the intrinsic total power of AGN indeed scales linearly with BH mass. 

\bibliographystyle{aipproc}   

\bibliography{sbs}

\IfFileExists{\jobname.bbl}{}
 {\typeout{}
  \typeout{******************************************}
  \typeout{** Please run "bibtex \jobname" to optain}
  \typeout{** the bibliography and then re-run LaTeX}
  \typeout{** twice to fix the references!}
  \typeout{******************************************}
  \typeout{}
 }

\end{document}